%% file: main.tex
\pdfoutput=1

\documentclass[10pt, conference, final]{IEEEtran}

\usepackage{cite}

\ifCLASSINFOpdf
\else
\fi

\usepackage[cmex10]{amsmath}

\usepackage{algpseudocode}

\PassOptionsToPackage{hyphens}{url}

\usepackage{multirow}
\usepackage[dvipsnames]{xcolor}
\usepackage{balance}
\usepackage{pifont}
\newcommand{\cmark}{\ding{51}} %
\newcommand{\xmark}{\ding{55}} %
\usepackage{tikz}

\newcommand*\circled[1]{\tikz[baseline=(char.base)]{
            \node[shape=circle,draw,inner sep=1pt,semithick] (char) {#1};}}

\newcommand*\dashcircled[1]{\tikz[baseline=(char.base)]{
            \node[shape=circle,draw,inner sep=1pt,dashed,semithick] (char) {#1};}}

\newcommand{\ignore}[1]{}

\newcommand{\remove}[1]{}
\newcommand{\mytodo}[1]{}
\newcommand{\added}[1]{#1}

\usepackage[printwatermark]{xwatermark}
\newwatermark[firstpage,color=black!100,angle=0,scale=1,xpos=0,ypos=129]{\small\textnormal{© 2020 IEEE. Personal use of this material is permitted.  Permission from IEEE must be obtained for all other uses, in any current or \\ future media, including reprinting/republishing this material for advertising or promotional purposes, creating new collective works,  \\ for resale or redistribution to servers or lists, or reuse of any copyrighted component  of this work in other works.}}
\newwatermark[allpages,color=black!100,angle=0,scale=1,xpos=0,ypos=-126]{\small 2020 50th Annual IEEE/IFIP International Conference on Dependable Systems and Networks (DSN) \\ Accepted version © 2020 IEEE. \url{https://doi.org/10.1109/DSN48063.2020.00043}}

\begin{document}
\title{The Impact of DNS Insecurity on Time}

\author{\IEEEauthorblockN{Philipp Jeitner\IEEEauthorrefmark{1}, Haya Shulman\IEEEauthorrefmark{1}, Michael Waidner\IEEEauthorrefmark{1}\IEEEauthorrefmark{2}}
\IEEEauthorblockA{\IEEEauthorrefmark{1}Technical University of Darmstadt, \IEEEauthorrefmark{2}Fraunhofer Institute for Secure Information Technology SIT}
}

\maketitle

\input{abstract}

\input{introduction}

\input{dns}

\input{cache-poison.tex}

\input{ntp-new.tex} %
\input{implementations} %
\input{server-side} %
\input{client-side} %

\input{mitigations.tex}

\input{conclusions}

\balance
\bibliographystyle{IEEEtran}
\bibliography{ntp,refs,NetSec}

\end{document}

%% file: abstract.tex
\begin{abstract}

\added{
We demonstrate the first practical off-path time shifting attacks against NTP as well as against Man-in-the-Middle (MitM) secure Chronos-enhanced NTP. Our attacks exploit the insecurity of DNS allowing us to redirect the NTP clients to attacker controlled servers. We perform large scale measurements of the attack surface in NTP clients and demonstrate the threats to NTP due to vulnerable DNS. 
}

\end{abstract}

%% file: introduction.tex
\section{Introduction}
Network Time Protocol (NTP) is one of the core Internet protocols meant to synchronise time on Internet systems. Due to its critical role in the Internet, NTP has a long history of attacks. %
Recently \cite{malhotra_attacking_2016} demonstrated a proof of concept attack against NTP that allows off-path attackers to shift time on victim clients. The idea of \cite{malhotra_attacking_2016} was to exploit overlapping IPv4 fragments, whereby a shifted time provided in attacker's fragments would overwrite the time provided by the real NTP server in a response sent to the NTP client. However, the limiting factors of the attack are: (1) the attack requires that the NTP servers agree to fragment the NTP responses to IP packets of 68 byte Maximum Transmission Unit (MTU) - \added{NTP servers do not fragment to such low MTU}, (2) the attacker must create and synchronise two spoofed fragments concurrently - \added{this is difficult to achieve even in lab conditions} and (3) the client must agree to shift time - since only a response from one NTP server is shifted while the rest return correct responses, the client would in most cases ignore it. Indeed, the authors found that only $0,0008\%$ of the NTP servers in the Internet could potentially be vulnerable to the attack.

{\bf Chronos.} Nevertheless, due to the critical role that NTP plays in the Internet there was an immediate followup work, which devised enhancements to NTP, called Chronos \cite{ntp:chronos}, to block that attack. Chronos is also on a standardisation track of the IETF \cite{ntp:draft}. Chronos leverages ideas from distributed computing on clock synchronisation in the presence of Byzantine adversaries and is designed to provide security even against strong Man-in-the-Middle (MitM) attackers and corrupted NTP servers. To enhance the security of NTP with Chronos only the NTP clients need to be modified. In contrast to ``plain'' NTP which queries few (typically up to 4) NTP servers, Chronos enhanced client queries time from multiple NTP servers, applies a secure algorithm for eliminating suspicious responses and averages the time over the valid responses. The authors demonstrate that in order to shift time on Chronos enhanced NTP client by 100ms a strong MitM attacker would need 20 years of effort.
{\em In this work, in addition to demonstrating practical off-path time-shifting attacks against plain NTP, we also show off-path attacks against Chronos enhanced NTP.}

{\bf Security in isolation.} Chronos \cite{ntp:chronos,ntp:draft} was designed to guarantee the security of NTP in isolation assuming an ideal environment and not taking into account other systems and protocols, like inter-domain routing, Domain Name System (DNS), Network Address Translation (NAT) devices. In the Internet, where multiple systems are running in concert, such a model does not encompass all the possible threats and attack vectors which can be exploited to subvert the security of a system. Indeed, as we demonstrate in this work, vulnerabilities in these systems can adversely affect the security of NTP. 

The authors in \cite{ntp:chronos} {\em implicitly} assume that the attacks on NTP are launched when the NTP client queries the NTP servers in the Internet for time. They model the attacker to be either a MitM that changes some responses to contain shifted time or an attacker that corrupts some of the NTP servers so that they provide invalid time. The security of Chronos is guaranteed if the majority of the queried NTP servers provide correct time. The idea is that all the responses are fed to a sophisticated algorithm which calculates time based on all the responses, the algorithm identifies and discards ``bad'' time and only considers responses with valid time. But, what happens when all, or a majority of, the responses are incorrect? In this case, Chronos will not be able to identify and discard bad responses. Luckily, even the most powerful MitM attacker cannot launch an attack to concurrently change the responses from {\em all or a majority of} the queried NTP servers - this is the idea underlying the design of Chronos.

{\bf The ``achilles heel'' of Chronos.} To generate a pool of NTP servers from which it then polls the time, the Chronos client periodically (every hour) queries DNS for domain {\footnotesize{\tt pool.ntp.org}}. The idea is to gradually collect hundreds of NTP servers, from which the Chronos client will be selecting the NTP servers at random and will be querying them for time. Then applying a sophisticating algorithm for filtering bad responses. We show that the achilles heel of Chronos are the multiple DNS queries.

\begin{figure}[t]
    \centering
    \includegraphics[width=0.45\textwidth]{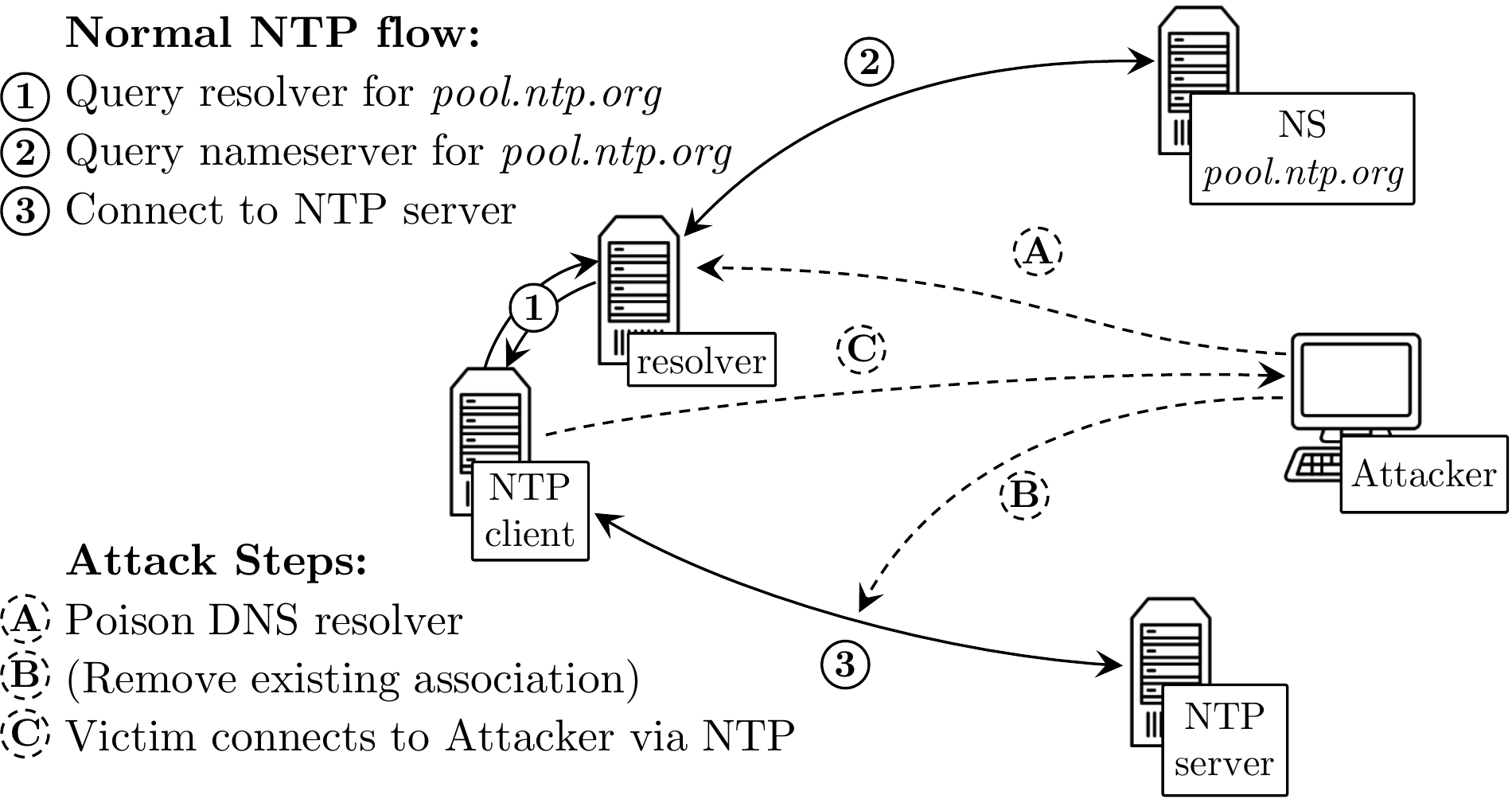}
    \vspace{-10pt}
    \caption{Attack overview}
    \label{fig:overview}
\vspace{-10pt}
\end{figure}

{\bf Our attack.} The idea of our attack is to leverage the cache poisoning vulnerability in DNS, and to inject malicious DNS records redirecting the DNS queries for {\footnotesize{\tt pool.ntp.org}} to an attacker controlled host. The attack consists of these phases (illustrated in Figure \ref{fig:overview}): the attacker needs to predict (or to trigger) a query from the victim DNS resolver to the nameserver {\footnotesize{\tt pool.ntp.org}} sent in step \circled{2} (we explain this in Section \ref{sec:attack}). The attacker performs DNS cache poisoning in step \dashcircled{A} (Section \ref{sc:dns}). The attacker causes the victim NTP client to drop the existing associations to real NTP servers in step \dashcircled{B} (Section \ref{sec:abuseratelimit}). Finally, the victim NTP client queries the attacker's NTP server and receives shifted time in step \dashcircled{C}.

These attacker controlled NTP servers provide incorrect time when receiving queries from NTP clients. Hence allowing the attacker to ensure that all (or a majority of) the responses from the malicious NTP servers are bad.

Our attack also applies against Chronos enhanced NTP client. In fact the multiple DNS requests issued by Chronos make the attack even easier to launch than against plain NTP. We demonstrate that these attacks can be launched even with an {\em off-path attacker}. Such attacks are even easier to launch with stronger attackers, such as those performing BGP prefix hijacking \cite{DBLP:conf/sigcomm/BallaniFZ07,DBLP:journals/pieee/ButlerFMR10,DBLP:journals/comsur/HustonRA11,dyn08,Cowie10}.

\added{
{\bf Countermeasures.} DNSSEC [RFC4033-RFC4035] would prevent the attacks, however, our measurements show that most NTP domains are not signed and most DNS resolvers do not perform validation. Only one domain ({\footnotesize{\tt time.cloudflare.com}}) is signed with DNSSEC, and only between 19.14\% and 28.94\% of the clients (depending on the geolocation) validate DNSSEC.
}

{\bf Ethical considerations.} Our attacks were tested against remote networks reliably, yet were ethically compliant. We measured and evaluated vulnerabilities in the DNS caches of the subjects of our study and measured which DNS caches are used by NTP servers on those networks, but did not hijack their traffic nor Internet resources and neither did we place incorrect DNS records for Internet domains that are not under our control in the caches of our subjects. Specifically, to avoid harming Internet customers and domains, we set up victim domains, which were used by us for evaluating the attacks, and evaluating the vulnerabilities. This ensured that the ``victim'' networks would not use the spoofed records for any ``practical'' purpose. We kept an extremely low attack volume and did not generate excessive traffic to avoid any interference with the operations of those services or load on them.

Our research shows that the NTP ecosystem is vulnerable to {\em practical} off-path cache poisoning attacks and worse, that the recent proposal Chronos for improving NTP security is even more susceptible to our attacks than the plain NTP. We are disclosing the vulnerabilities along with recommendations for countermeasures.

{\bf Our contributions.} %

$\bullet$ In this work we improve over the attacks presented in \cite{malhotra_attacking_2016} and demonstrate the first off-path attacks against Chronos-enhanced NTP that was proven secure even against the strong MitM attackers. 

$\bullet$ {\em New attack methodologies.} We develop new methods to launch the attacks when the attacker does not control the timing of the queries and the timing is difficult to predict: (1) recurring IP defragmentation cache poisoning until the attack succeeds, (2) utilising other systems using the same DNS resolver for initiating the attack and issuing the query. 

\ignore{
leverage the insecurity of DNS. Specifically, we poison the DNS caches used by NTP clients, hence redirecting the clients to ``malicious'' NTP servers all of which provide bad time in responses. 

$\bullet$ {\em On a technical level, we designed attacks against plain NTP and against Chronos enhanced NTP clients.} Our attacks improve over the off-path NTP attacks of \cite{malhotra_attacking_2016} and are more efficient and effective, applying to a large number of NTP servers and clients. %
The authors provided a proof of concept attack but could not evaluate it against Internet servers since it requires reducing the MTU of nameservers to 68 bytes and requires synchronising two overlapping fragments. The attack also requires that the client accepts the response with the bad time - in reality clients query a number of NTP servers and would ignore servers that provide bad time. In contrast, our attack is more practical: depending on the DNS response even a 500-byte fragment may suffice to launch the attack, and only a single fragment from the attacker is needed. Additionally, since all the NTP queries are sent to attacker's NTP servers, the attacker ensures that all the responses serve identical time, causing the NTP client (as well as Chronos enhanced NTP client) to deterministically accept it. Furthermore, we demonstrate that the recent proposal \cite{ntp:chronos} for improving NTP security against MitM attackers is vulnerable to our off-path attacks. %
}
$\bullet$ {\em Adaptation of general cache poisoning techniques for attacks against NTP.} For our cache poisoning attacks we apply defragmentation cache poisoning attacks presented in \cite{cns:frag:dns}, and adapt them for attacks against NTP. The idea is to inject a spoofed second fragment that is reassembled with the first fragment provided by the real nameserver. This allows to match the challenge response parameters without having to guess them because they are in the first fragment, while injecting malicious payload in the second fragment. We provide evaluation of these techniques against the NTP ecosystem. During run-time the NTP client already has associations to other NTP servers and hence will not use the injected record with the new NTP server. We develop a technique which causes the NTP client to break established associations to NTP servers and {\em pin} the client to the NTP server provided by the attacker. To do this, we exploit the rate limiting defence supported by the vast majority of NTP servers in the Internet.

$\bullet$ {\em We measure the attack surface introduced by our techniques on NTP ecosystem.} We perform extensive measurements to identify the NTP clients that are vulnerable to our off-path attacks. This essentially means NTP clients that use DNS resolvers that are vulnerable to cache poisoning attacks.

$\bullet$ {\em Conceptually our work demonstrates the ``weakest link'' problem in the Internet.} Our work provides insights to the gap between theoretical proofs of security done in isolated environments with assumptions which, unfortunately, do not hold in practice. This exposes the systems, even when proven secure, to unexpected attacks. As we show in this work, NTP heavily depends on DNS for its security, hence security of NTP cannot be established in isolation from DNS. Unfortunately, as is demonstrated by the research community as well as attacks that were detected in the Internet, DNS is extremely vulnerable to DNS cache poisoning.

$\bullet$ We provide more information about this project at http://ntp-attack.sit.fraunhofer.de.

{\bf Organisation.} We review Related Work in Section \ref{sc:works}. In Section \ref{sc:dns} we explain how to exploit IP fragmentation for launching DNS cache poisoning attacks and discuss the hurdles which the attacker needs to overcome. In Section \ref{sec:attack} we explain the adaptations of the attack for attacking NTP: specifically, triggering the DNS request and then causing the victim NTP client to use the NTP server provided by the attacker. In Section \ref{sec:impl} we evaluate the attack against different implementations of NTP clients. In Sections \ref{sec:measurements} and \ref{sec:ntpdns_resolvers} we provide our measurements of the attack surface of DNS and NTP servers as well as the DNS resolvers. In Section \ref{sec:mitigations} we provide recommendations for countermeasures. We conclude this work in Section \ref{sec:conclusion}.

%% file: dns.tex
\section{Related work}\label{sc:works}
\label{sec:relwork}

{\bf NTP Security.} Network Time Protocol (NTP) is one of the Internet's oldest protocols that was designed in the 80s. NTP has a long history of attacks. NTP was typically exploited as a reflector to launch Distributed Denial of Service (DDoS) attacks against victims in the Internet, \cite{czyz2014taming,Amplification:Hell}.

NTP can also be attacked via non-standard queries and via Config interface, \cite{ntp:kiayias_security_2017}. Although the recent patches introduced to NTPv4.2.8p9 were meant to close the vulnerabilities, however, our measurements performed in this work indicate that at least 5\% of NTP servers still have open Config.

The researchers also investigated the implications of shifting time on security of computer systems and applications and integrity of timing information \cite{mills2016computer,rfc7384}. Additionally, evaluations were performed of on-path (Man-in-the-Middle) attacks for shifting time on NTP clients, along with analysis of the potential impact of the attacks, \cite{selvi2014bypassing,selvi2015breaking}.

Recently \cite{malhotra_attacking_2016} demonstrated that it is theoretically possible to shift time with an off-path attacker, however, their attack assumes that the attacker can reduce the responses from NTP servers to a very low MTU of MTU=68, which is unrealistic, and hence the authors could only present a proof of concept attack in a lab setup. Furthermore, the assumption of \cite{malhotra_attacking_2016} that the victim NTP clients support a specific fragments' reassembly strategy could not be verified in practice: it requires generating two fragments that overlap with the NTP response of the NTP server and overwrite a few locations in the NTP packet. The attack also requires that the client accepts the response with the bad time - in practice clients query a number of NTP servers and ignore servers that provide bad time. 

Our work is similar to \cite{malhotra_attacking_2016} as we also apply fragmentation in order to cause the victim NTP client accept an incorrect time in a response to its NTP request. However, in contrast to \cite{malhotra_attacking_2016} we do not attack the time-responses from NTP server, but we target the DNS responses from the nameserver. Our attack is also more practical in that it does not require sending two overlapping fragments - which is difficult to match correctly in practice, but we need to send a single (second) fragment which contains the malicious payload and redirects the victim NTP client to query the ``wrong'' NTP servers. Specifically, instead of providing an incorrect time directly, we first redirect the victim NTP client to ``query'' the attacker's NTP servers, which in turn provide shifted time. Our attacks are also more effective: we do not attack a response from a single NTP server, but provide shifted responses from {\em all} the NTP servers that the client queried. This allows us to ensure that the client definitely accepts the shifted time.

In a recent work, NTP 'prime-the-pump' \cite{malhotra_attacking_2016-1} used a Kiss of Death (KoD) packet to the victim NTP client, to cause it to slow down requests to that NTP server. In contrast, we abuse the rate limiting mechanism of NTP to cause the client to ``forget'' the server and replace it with an attacker controlled one via a malicious DNS record that we inject into a resolver's cache that the victim NTP client is using.

Although NTP support cryptographic authentication, \cite{dowling2016authenticated} in practice NTP is generally not cryptographically protected \cite{malhotra_attacking_2016}.

There were also proposals for probing time from distributed locations to provide security against MitM attackers, \cite{mizrahi2012slave,shpiner2013multi}. These proposals are not deployed in the Internet due to the significant changes that adoption requires. Recently, \cite{ntp:chronos} proposed to generate NTP-servers redundancy by collecting a sufficiently large set of NTP servers through multiple DNS queries. Chronos was also proven to provide security against even strong MitM attackers. Nevertheless, as we show, exactly the multiple DNS queries make Chronos even more susceptible to our off-path attacks than the plain NTP.

\ignore{
{\bf DNS Security.} DNS cache poisoning allows off-path attackers to gain MitM capabilities for the victim domain - effectively to hijack the target domain in some victim DNS resolver. Assume an off-path attacker poisons the cache of a target resolver by injecting a record mapping the victim domain {\tt vict.im} to an IP address controlled by an attacker, say {\tt 6.6.6.6} instead of the actual IP address {\tt 123.0.0.53} of the real nameserver authoritative for {\tt vict.im}. As a result, any query for a resource within the target domain {\tt vict.im} will redirect the traffic to the attacker's host {\tt 6.6.6.6} instead of {\tt 123.0.0.53}. This allows the attacker to intercept any communication to the victim domain, such as web, email, FTP. If the attacker further relays the traffic to the legitimate destination or generates the responses that look correct himself - the attack will not be detected.

Such attacks were known since the 90s, and evaluated in the lab were considered mostly theoretical \cite{Vixie95,Bernstein:DNS}.
}

{\bf DNS Security.} DNS cache poisoning attacks were known since the 90s, and evaluated in the lab were considered mostly theoretical \cite{Vixie95,Bernstein:DNS}. In 2008 Kaminsky \cite{kaminsky:dns} demonstrated the first practical DNS cache poisoning attack. To enforce a query the attacker would request a random subdomain of the target victim domain. In the spoofed responses the attacker adds a new malicious nameserver for the victim domain, which is added to the cache if the attack succeeds. 
Following Kaminsky attack many systems were patched to support the best practices in [RFC5452], \cite{hubert2009measures}. The main patches were challenge-response authentication based on randomisation of source port and transaction identifier (TXID) values in DNS packets, each of which is 16 bits.

After deployment of source and TXID randomisation off-path attacks became a theoretical threat. To launch successful cache poisoning attacks, the attackers have to be MitM, i.e., have to be able to see the queries. %

However, in 2011 \cite{cns:frag:dns} demonstrated practical off-path DNS cache poisoning attacks that use fragmented IPv4 packets. The idea is to inject a spoofed fragment into IP defragmentation cache that is to be reassembled with the real fragment from the nameserver, and so to inject malicious payload into the DNS response. Fragmentation is a popularly exploited attack vector for attacks against different systems and protocols \cite{gilad2014off,shulman2015towards,malhotra_attacking_2016,brandt2018domain}.

%% file: cache-poison.tex
\section{DNS Poisoning via Fragments Replacement} \label{sc:dns}

The basic component that we use to launch a DNS cache poisoning attack is to manipulate a DNS response sent from the nameserver to the victim resolver via replacement of IPv4 fragments. We exploit the replacement of fragments to overwrite part of the payload in a real DNS response from the nameserver with malicious values. This allows the off-path attacker to inject malicious DNS records into the cache of a victim DNS resolver without having to guess the values in the challenge-response fields (port, TXID), \cite{rfc5452,herzberg2012antidotes} of the DNS response (those are transmitted in the first fragment). 

We next list the components needed for launching a DNS cache poisoning attack.

\subsubsection{IP fragmentation} DNS responses typically do not exceed 1500 bytes and hence do not fragment. To cause the nameserver to send a fragmented DNS response the attacker sends to the nameserver a \textit{ICMP Destination Unreachable Fragmentation Needed} error message (type 3, code 4) with DF bit set. The error tells the nameserver that the Maximum Transmission Unit (MTU) to the destination is smaller. Upon receiving the ICMP error the nameserver registers the MTU indicated in the ICMP error and sends the DNS response in IP fragments that do not exceed the MTU indicated in the ICMP error.

\subsubsection{IP defragmentation cache poisoning} Assume that the nameserver sends a response in two fragments. The attacker generates a second fragment that is identical to the second fragment of the nameserver except for the new malicious records - IP addresses or nameserver records - that the attacker adds. These records map the NTP servers to hosts controlled by the attacker.

To cause the victim resolver to reassemble the spoofed second fragment with the first fragment of the nameserver the attacker has to ensure that both fragments contain the same source and destination IP addresses, the same IPID value\footnote{IPID, [RFC6864], is used to identify all fragments belonging to the same original IP packet. IPID values can often be predicted.}, and that both are fragments (i.e., {\tt more fragments} flag is set in the first fragment and is zero in the second (last) fragment). The fragments are then reassembled based on the offset values.

To predict the IPID that will be assigned by the nameserver to the response sent to the victim DNS resolver, the attacker sends a few queries to the nameserver. This allows to measure the current IPID value and the rate at which the value is incremented. The attacker uses that information to extrapolate the value that will be used in a DNS response that will be sent by the nameserver to the victim resolver. We use the same IPID prediction techniques as those developed earlier \cite{cns:frag:dns,frag:vulnerable:journal}. This will be the IPID value that the attacker will set in its spoofed second fragment sent to the victim DNS resolver. Notice that if the rate at which the nameserver receives queries is high, and IPID increments are high, the attacker can send multiple fragments, to cover a set of possible IPID values. Most stringent operating systems, such as Windows and patched Linux, allow 100 and 64 identical fragments respectively, each with a different IPID value.

\subsubsection{Calculating UDP checksum} If the spoofed second IP fragment is correctly constructed, it will be reassembled with the first IP fragment of the nameserver and the UDP payload will be passed on to the transport layer and then to the DNS software. These layers will perform subsequent checks, including the UDP checksum, the packet length, and correctness of the DNS records. The next challenge is matching the UDP checksum.

Since the off-path attacker modifies part of the payload of the original IP packet sent by the nameserver it has to ensure the UDP checksum matches the one of the original IP packet. The UDP checksum value resides in the first fragment in UDP header and cannot be altered by the attacker. The attacker needs to ``fix'' the checksum by adjusting some other two bytes in the second fragment.

UDP checksum calculation is performed as follows: a ones' complement sum is calculated on all the 16-bit values in the packet and then the ones' complement (i.e., invert all bits) is taken of that value to the checksum field.
This means that with knowledge about the original second fragment $f_2$, the attacker
can ensure a matching checksum by crafting the modified fragment $f_2'$ so that
the ones' complement sum $sum_1(f_2') = sum_1(f_2)$. This is possible for example by
measuring the change in ones' complement sum between $f_2$ and the preliminary modified fragment ${f_2}^*$ and subtracting it from unimportant 16-bit values, therefore creating $f_2' = {f_2}^* - (sum_1({f_2}^*) - sum_1(f_2))$.

%% file: ntp-new.tex
\section{Redirection to Attacker's NTP Servers}
\label{sec:attack}
In this section we explain under what conditions the attacker can trigger DNS requests for launching the DNS cache poisoning attack and injecting a spoofed DNS record mapping to a malicious NTP host. Then, how the attacker can cause the client to use the malicious record. We explain that at Boot time the victim NTP client will use the injected record directly since it does not have any other associations. At run-time, we demonstrate how to exploit rate limiting supported by NTP servers to cause the NTP client to break its associations to NTP servers in order to use the newly injected record of the attacker.

\subsection{Attacking NTP during boot-time}
\label{sec:boottime}
\begin{figure}[t]
    \centering
    \includegraphics[width=0.45\textwidth]{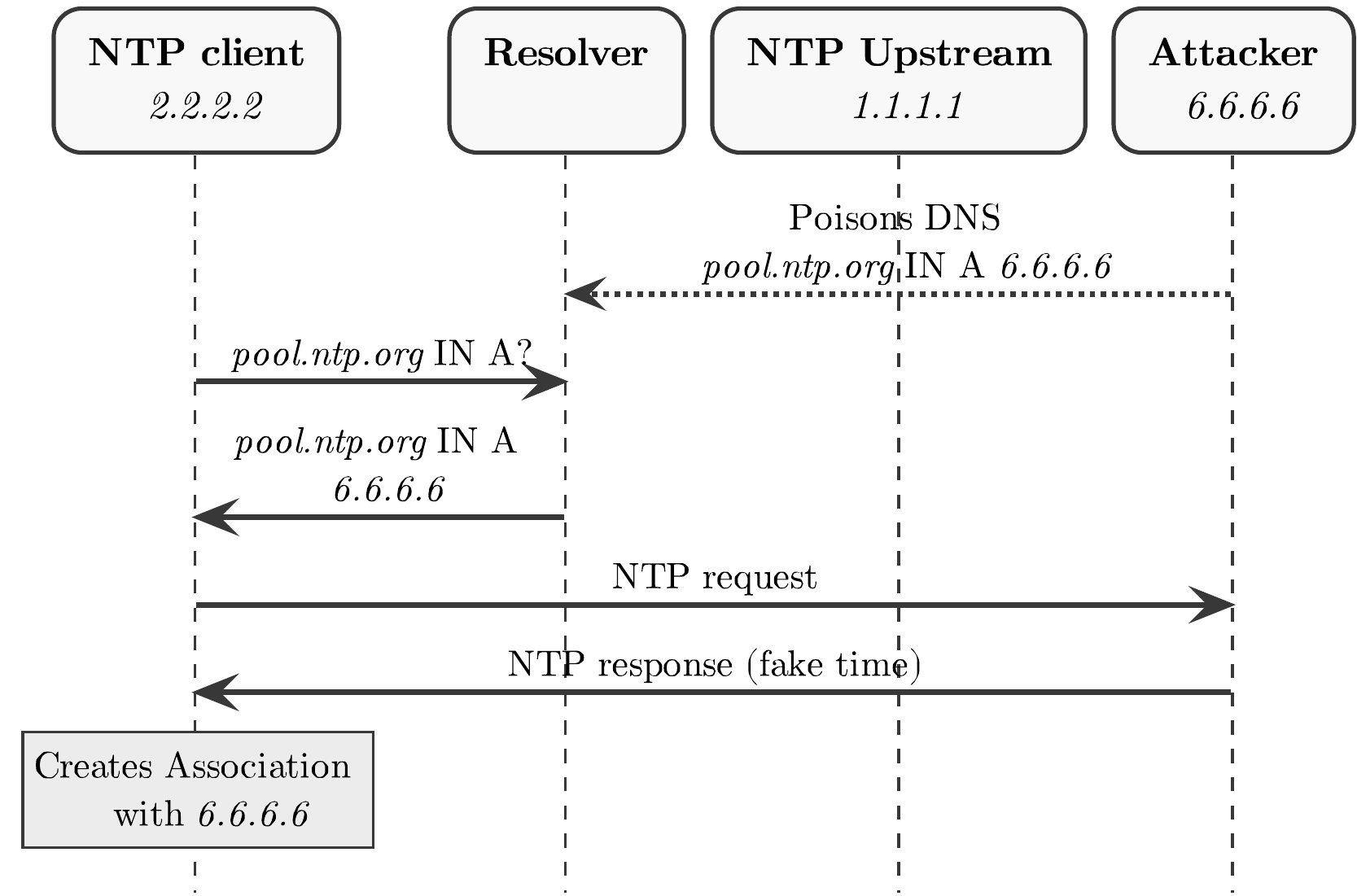}
    \vspace{-10pt}
    \caption{Boot-time attack on NTP.}
    \label{fig:boot-time-attack}
\vspace{-10pt}
\end{figure}

The attack on NTP during boot-time is shown in Figure~\ref{fig:boot-time-attack}. After startup, the NTP client immediately queries it's DNS resolver to find NTP servers to associate to. If the attacker was able to poison the DNS resolver's cache before, the resolver will return the IP address ({\it 6.6.6.6}) controlled by the attacker in the DNS response and the NTP client associates with the NTP server at that address, thus takes time from the attacker.

\added{
There are three ways to initiate the cache poisoning attack at boot-time: (1) by predicting the time of the query via side channels, (2) by causing other systems that use the same DNS resolver, such as Email or open resolvers, to issue the query, or (3) by periodically planting the spoofed (second) fragment every 30 seconds in the IP defragmentation cache of the victim resolver until the NTP client issues the query and the response is reassembled with the spoofed fragment waiting in the defragmentation cache. We measured frag. reassembly timeout of 60 to 120 seconds on Windows distributions and 30 seconds on Linux; [RFC2460] specifies the reassembly timeout of 60 seconds. This approach requires a low attack volume which can be completed with only one low bandwidth attacking host. Furthermore, the TTL of {\footnotesize{\tt pool.ntp.org}} A record is only 150 sec. Namely, the resolver will frequently issue queries for {\footnotesize{\tt pool.ntp.org}}, at most every 150 sec. This means that the attacker needs to maintain the spoofed fragment in the IP defrag. cache for at most 150 sec, which in the worst case requires 150/30 = 5 spoofed (second) fragments per attack. Referring to option (2), in Section~\ref{sec:sharedresolvers} we show that DNS resolvers are indeed often shared between different systems, so that the system used to poison the cache can be different from the system targeted by the attack.
}

\subsection{Attacking NTP during run-time}
\label{sec:abuseratelimit}
\begin{figure}[t]
    \centering
    \includegraphics[width=0.47\textwidth]{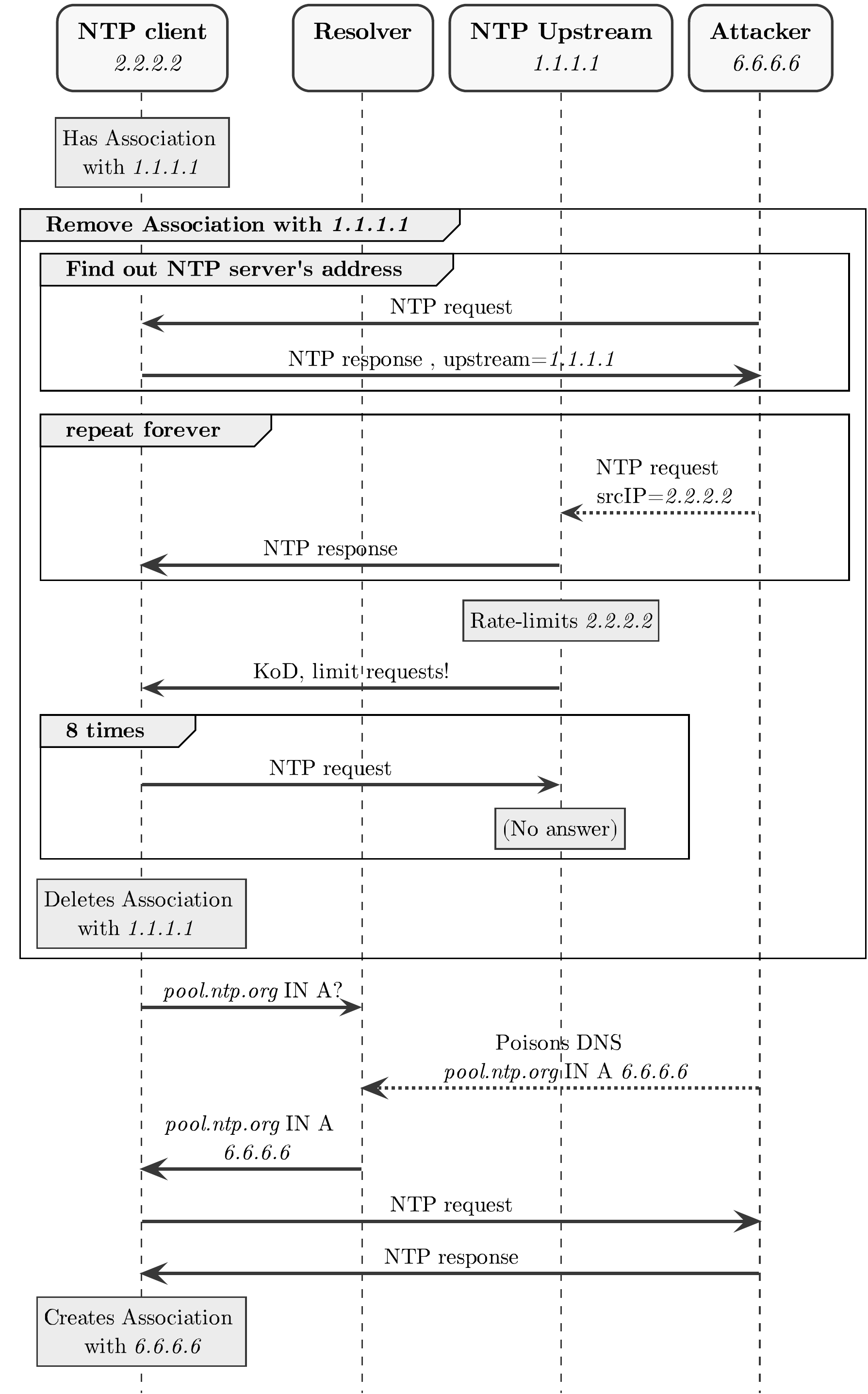}
    \vspace{-10pt}
    \caption{Run-time attack on NTP.}
    \label{fig:run-time-attack}
\vspace{-10pt}
\end{figure}

During run-time, the NTP client already has a list of associations with NTP servers which addresses are resolved. This means that poisoning the resolver's cache during run-time will not directly lead the NTP client to associate to the attacker's NTP server. However, we show how the attacker can still trigger a DNS query and make the client connect to the attacker controlled server. This attack is shown in Figure \ref{fig:run-time-attack}. First, the attacker needs to remove the existing association to the server at $1.1.1.1$, which will cause the client to replace the server with another one by querying the DNS resolver and therefore, trigger the DNS query.

\subsubsection{Removing the existing association}

To remove the existing association with the NTP server at $1.1.1.1$, the attacker needs to disrupt the communication between the NTP client and server. This way, the NTP server will appear unreachable to the NTP client, and therefore lead the client to replace the unreachable server via DNS. While the easiest way to disrupt this communication would be a Denial-of-Service attack on the NTP server, this method may require a huge amount of resources and won't go unnoticed for long. Instead, we show how to facilitate the rate-limiting mechanism used by many NTP servers to make the NTP server appear unreachable to the client even though it is not.

\subsubsection{Exploiting rate limiting of NTP}
Many public NTP servers in the internet enable rate-limiting to limit the server load caused by defective NTP clients or the impact on reflection attacks. Similar to DNS, NTP servers can be used as reflectors of Denial-of-Service attacks which use spoofed source IP addresses to hide the attacker's real address. The rate-limiting mechanism in the default implementation of NTP works by monitoring the amount of NTP queries from a single IP and denies service, ie. stop answering requests for some time, if the time-span between two NTP queries is too low. In previous works, Malhotra et al. \cite{malhotra_attacking_2016} already showed how this mechanism can be used to force an NTP client to take time from another NTP server by triggering 'Kiss-o-death' packets, which are NTP responses generated to indicate that the client should reduce it's query interval and are generated just before the server starts rate-limiting the NTP client. Instead of relying on the client to reduce it's query interval, our attack relies on the server-side rate-limiting which will cause the server to stop replying to NTP requests at all. In Figure \ref{fig:run-time-attack}, this is shown as the attacker keeps sending spoofed NTP requests with the NTP client's source IP address so that the server finally rate-limits the client and stops answering any subsequent requests. After that, the NTP client will keep trying to query the NTP server several times after it gives up and marks the association as unusable. Depending on the client, it will then search for a new server immediately via DNS or switch to another one if it uses multiple associations simultaneously. Consequently, to succeed with the attack the attacker may need to remove multiple existing associations of the NTP client using the method presented here.

To find out the address of the NTP server(s) used by the client, which is needed to create the spoofed NTP requests, the attacker has multiple options:

\paragraph{Generating a list of IP addresses} The attacker queries the DNS system for the domain name configured in the NTP client and create a list of possible upstream NTP server addresses. For the {\footnotesize{\tt pool.ntp.org}} domain this list consists of 2000 to 3000 servers, which is small enough for the attacker to successfully disrupt existing associations with any of these servers by abusing NTP's rate limiting mechanism.

\paragraph{NTP Upstream server address leakage} If the target NTP client is also acting as a NTP server (eg. NTPd at default configuration), the attacker can use the `id' field in the NTP response to identify the currently used upstream server's IP address. In this case, the attacker only learns one server address at a time and needs to wait until the upstream server is changed until it learns the next server's IP address. This is the variant shown in Figure \ref{fig:run-time-attack}.

\paragraph{Querying the servers configuration interface} Some servers running NTPd expose the NTP configuration interface into the Internet which can be used to learn both, the configured DNS hostnames as well as all currently used upstream server IP addresses. Furthermore, \cite{ntp:kiayias_security_2017} showed that this interface can also be exploited for other attacks. Nevertheless, we found that 5.3\% of the NTP servers in the {\footnotesize{\tt pool.ntp.org}} pool still respond to configuration queries during our measurements in section \ref{sec:ntpdns_ratelimit_upstream}.

\added{
\subsubsection{Conducting the cache poisoning attack}
To place the malicious DNS record in the vicitm resolver's cache, the attacker can employ the same methods as discussed in Section~\ref{sec:boottime}. Predicting the timing of the query in this case is easier as the query is triggered by the attacker himself when the client re-queries its own resolver after it has deleted the existing association. However, utilising other systems to place the malicious record in the cache might still be easier as those systems might allow the attacker to avoid or override cached records and evict records from the cache by selecting a custom query domain which is not possible with NTP itself as the query domain cannot by influenced by the attacker.}

%% file: implementations.tex
\section{Evaluation of Plain NTP Clients}
\label{sec:impl}
In this section we evaluate which NTP clients are vulnerable to boot-time attacks and which are vulnerable to run-time attacks. We then provide an analysis of successful time-shifting attacks by causing the victim NTP client to query the malicious NTP server of the attacker and accept from it a shifted time.

\subsection{DNS Lookup Behaviour of NTP Implementations}
\label{sec:impl_dns}

We evaluate applicability of our attacks to popular NTP clients. Our evaluation is performed in both attack scenarios presented in Section \ref{sec:attack}: at boot-time and at run-time. The evaluation is based on analysis of DNS lookup behaviour of the NTP implementations using: source-code and lab evaluation of the run-time attacks against the NTP clients with DNS poisoning. %
The clients support either NTP or Simple NTP (SNTP). The main difference between those protocols is that SNTP uses a single NTP server whereas NTP can use multiple NTP servers. In the second column (from left) in Table \ref{tab:ntpclients} we list the different NTP client implementations. In the third column we list the distribution of different NTP clients and NTP servers in {\footnotesize{\tt pool.ntp.org}}. In the last two columns we report which attack setup (boot-time or run-time) applies to which NTP client implementation.
\added{As listed, all NTP implementations are vulnerable to a boot-time-attack and 4 NTP implementations are vulnerable to a run-time attack. These 4 out of 7 NTP implementations make up at least 45\% of clients of {\footnotesize{\tt pool.ntp.org}}.}

\begin{table}[t]
\begin{center}
\scriptsize
\caption{Attack scenarios for popular NTP clients.}
\label{tab:ntpclients}
\begin{tabular}{ |c|c|r|c|c| }
    \hline
          & Client            & \emph{pool.ntp.org} & \multicolumn{2}{c|}{ Attacks } \\
    \cline{4-5}
          &                   & Usage \cite{rytilahti_masters_2018} & boot-time & run-time \\
    \hline
    \multirow{3}{*}{ \rotatebox[origin=c]{90}{NTP}}
          & NTPd              & $26.4\%$ & \cmark & \cmark \\
          & openntpd          & $4.4\%$  & \cmark & \xmark \\
          & chrony            & $4.8\%$  & \cmark & \cmark \\
    \hline
    \multirow{4}{*}{ \rotatebox[origin=c]{90}{SNTP}}
          & ntpdate           & $20.0\%$ & \cmark & n/a    \\
          
          & Android           & $14.0\%$ & \cmark & \cmark \\
          
          & ntpclient         & $1.2\%$  & \cmark & \xmark \\
          & systemd           & \emph{not listed} & \cmark & \cmark \\
    
    \hline
\end{tabular}
\end{center}
\vspace{-20pt}
\end{table}

\subsubsection{Boot-time attacks}

Our analysis shows that all types of NTP clients are vulnerable to cache poisoning at boot-time when performing DNS lookups, as there is no viable mitigation mechanism against the attack itself. Some clients enforce limits to how much or how fast time can be shifted from the local system clock, however, these limits are often not used in the default configuration or are explicitly not enforced at boot-time, because of the assumption that the system clock may be way off at the time the system starts because of dead Real-time-clock batteries etc. The only exception to this we saw for traditional NTP clients is the openntpd client, which has the option to partially authenticate the time set via using the {\tt Date}-header of a HTTP response gathered using a TLS-protected connection to a configurable webserver. This is not enabled by default.

\subsubsection{Run-time attacks}

Next, we show and evaluate the vulnerability against run-time attacks of those NTP implementations. Using documentation and lab tests, we discovered that only 4 of our clients are vulnerable to run-time attacks: NTPd, chrony, Android and systemd-timesyncd. openntpd and ntpclient do not support DNS queries during run-time at all, so hindering communication with the used servers will just disable time synchronisation until the client is restarted. A special case is ntpdate, which is a command-line utility which only synchronises time once and then exits, so that the run-time attack does not apply here as well. However, this utility is often used as part of a regularly run cronjob, so boot-time attacks against this client can be done any time the program is invoked. We practically evaluated all potentially vulnerable clients except Android, which was discarded from the list because we found no device which actually used the built-in NTP client, all available devices used the mobile network to synchronise time instead. However, from source code reviews of code, we can defer that since the built-in NTP client is always invoked by hostname\footnote{Source code: \url{https://android.googlesource.com/platform/frameworks/base/+/master/core/java/android/util/NtpTrustedTime.java}}, DNS lookups must be triggered every NTP query if not answered from a local DNS cache.

For the other 3 clients we performed DNS poisoning attacks using a DNS resolver reconfigured after the clients had done their initial boot-time DNS lookups. We then used a custom application to send spoofed NTP mode 3 queries to all 4 of our labs NTP servers to trigger the rate-limiting mechanism and stop communication with our client. After some time, all 3 clients re-queried the DNS resolver to find new servers and switched over to our attacker-NTP server which provided time shifted by -500 seconds. All 3 clients started to adjust their system clocks after some time, the exact timing values are listed in Table \ref{tab:labattackdurations}. \added{For ntpd, we did this evaluation two times, one time assuming the attacker knows the upstream NTP servers' addresses in advance (Scenario $P_1$) and one time where the attacker discovers these addresses one-at-a-time by querying the client via mode NTP 3 queries (Scenario $P_2$).}

\begin{table}[t]
\begin{center}
\scriptsize
\caption{\added{Run-time attack duration against different clients}}
\label{tab:labattackdurations}
\begin{tabular}{ |r|c|l| }
    \hline
          Client    & Scenario & Attack duration \\
    \hline
          NTPd      & $P_2$    & 47 minutes \\
          NTPd      & $P_1$    & 17 minutes \\
          openntpd  & $P_1$    & 84 minutes \\
          chrony    & $P_1$    & 57 minutes \\
    \hline
\end{tabular}
\end{center}
\vspace{-20pt}
\end{table}

\subsection{Probability Analysis of Run-time Attack}
\label{sec:impl_prob}

To successfully shift time on a NTP client, an attacker needs to replace a majority of NTP 
associations this client uses. In this section we give a probabilistic analysis of whether
an attacker is able to do this on a running NTP client, depending on his knowledge of the
used upstream server's IP addresses.

\subsubsection{Scenario 1}

In this scenario, we have a client which has a list of upstream servers, which we remove one-after-another by attacking the connection between client and server. This is the case if the attacker cannot know the upstream servers' addresses upfront but instead discovers it via querying the client as shown in Figure \ref{fig:run-time-attack}.  We assume that the client has chosen it's upstream servers randomly from the NTP pool, so the probability that each of the servers in the list does rate limiting is $p_{rate}$. Therefore the probability of successfully removing $n$ servers by exploiting the server's rate-limiting mechanism is
$P_1(n) = p_{rate} ^ n$

\subsubsection{Scenario 2}

In the second scenario, we assume that the attacker has knowledge about the clients upstream servers upfront and therefore can choose the servers which shall be removed from the list. This can be achieved when the server leaks configuration information or simply by attacking potential connections to all servers in {\footnotesize{\tt pool.ntp.org}}. Assuming a server list of size $m$, we calculate the probability of a successful attack as the probability that $n$ or more servers out of $m$ do rate-limiting:

$$P_2(m,n) = \sum_{i=n}^{m} \overbrace{\underbrace{{m \choose i}}_{\makebox[0pt]{\textrm{\tiny{Number of possibilities to choose}}} \atop \makebox[0pt]{\textrm{\tiny{i out of m servers}} }} \cdot p_{rate}^i \cdot p_{\overline{rate}}^{m-i}}^{\makebox[0pt]{\textrm{\tiny{Probability that exactly i out of m servers do rate limiting}}}}$$

If $n=m$, this is the same as $p_{rate} ^ n$.
As said, an attacker needs to replace a majority of the used upstream servers to successfully shift time, therefore we have $n \geq \left\lceil\frac{m}{2}\right\rceil$. For $p_{rate} = 38\%$ as indicated by a scan of all servers in {\footnotesize{\tt pool.ntp.org}} described in Section \ref{sec:ntpdns_ratelimit_upstream} we list the probabilities to conduct a successful attack in Table \ref{tab:ntp_propabilies}. 

\begin{table}[t]
\begin{center}
\scriptsize
\caption{\added{Probabilities that an NTP client is in a vulnerable state depending on the number of used associations $m$.}}
\label{tab:ntp_propabilies}
\begin{tabular}{ | l | l | l | l | }
    \hline
    $m$ & $n=\max(\left\lceil\frac{m}{2}\right\rceil, m-2)$ & $P_1(n)$ & $P_2(m,n)$ \\
    \hline
    1   &   1   &   38.0\%   &   38.0\%   \\
    2   &   2   &   14.4\%   &   14.4\%   \\
    3   &   2   &   14.4\%   &   32.4\%   \\
    4   &   3   &   5.5\%    &   15.7\%   \\
    5   &   3   &   5.5\%    &   28.4\%   \\
    {\bf 6}   &   {\bf 4}   &   {\bf 2.1\%}    &  {\bf  15.3\% }   \\
    7   &   5   &   0.8\%    &   7.8\%    \\
    8   &   6   &   0.3\%    &   3.9\%    \\
    9   &   7   &   0.1\%    &   1.8\%    \\
    \hline
\end{tabular}
\end{center}
\vspace{-15pt}
\end{table}

\added{
\subsubsection{Common client configurations}
What probabilistic scenario applies to what client depends in its configuration and implementation behaviour. For example, ntpd in most default configurations is configured to have 4 persistent pool associations which are used to create server associations using DNS lookups when needed. The maximum long-time number of associations is 10 ({\tt \small NTP\_MAXCLOCK}), so together with the pool-associations, a default ntpd client will have $m=6$ NTP upstream servers and will only query for new associations during run-time if this number is reduced below 3 ({\tt \small NTP\_MINCLOCK}). This means the attacker needs to remove $n = m-2 = 4$ servers to trigger a DNS query. Notably, neither {\tt \small NTP\_MAXCLOCK} nor {\tt \small NTP\_MINCLOCK} can be configured, the only variable which can be changed is the number of configured pool associations, which effectively \emph{lowers} the amount of active server associations as they count towards the {\tt \small NTP\_MAXCLOCK} limit. The probability for a run-time attack against ntpd is therefore $P_1(4)$ or $P_2(6, 4)$.

The default configuration for systemd-timesyncd includes only a single ntp pool domain. As systemd-timesyncd is a SNTP client, it holds only a single association to one NTP server but caches the list of servers from the last DNS query, which by default contains 3 more server addresses additional to the one used. As these servers will be queried before a DNS query is triggered, the attacker is required to attack associations to all of them, giving a run-time attack probability of $P_1(4)$, since all of these 4 servers need to be removed. If more domains are configured, {\tt systemd-timesyncd} will round-robin through all of them but since no responses are cached except for the first domain, this will still trigger a DNS query.
}

\section{Evaluation of Chronos NTP Client}
\label{sec:chronosdiscussion}

Chronos \cite{ntp:chronos} is a special NTP client proposed to prevent time-shifting attacks by sampling time
from a bigger set of NTP servers than traditional NTP implementations. It is also available as an internet standard draft  document \cite{I-D.schiff-ntp-chronos}. While Chronos uses a proven secure method for setting the time using a subset of a big pool of NTP nameservers, it is not stated exactly how this server pool is generated, which is important since the security guarantees of Chronos vanishes if the attacker is able to control more than $\frac{2}{3}$ of the NTP servers in the pool. The original proposal describes the pool generation works by querying the {\footnotesize{\tt pool.ntp.org}} domain every hour for 24 hours and use the union of all gathered IP-addresses as the server pool. Because the nameservers of {\footnotesize{\tt pool.ntp.org}} normally give 4 IP-addresses per DNS query, this results in a maximum of 96 servers. However, since the DNS namserver controls the servers Chronos considers during its  algorithm, an attacker attacking the DNS instead of NTP can feed many of its own IP-Addresses into Chronos' server pool, thereby overwhelming the honest servers. In the following we point out two weaknesses of the proposed server-pool generation mechanism and after that, describe a potential DNS poisoning attack on Chronos.

\begin{figure}
    \centering
    \includegraphics[width=0.45\textwidth]{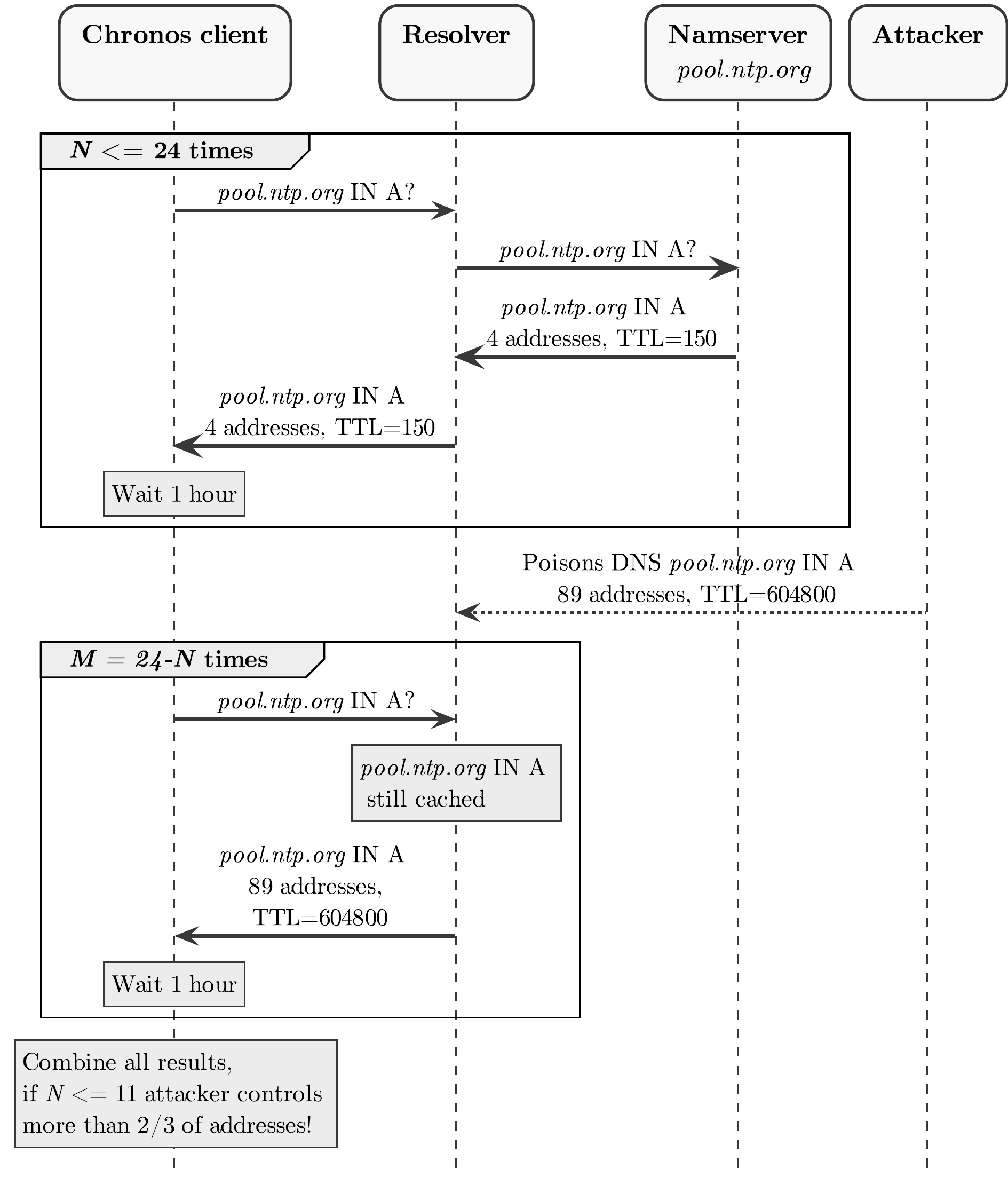}
    \vspace{-10pt}
    \caption{DNS poisoning attack on chronos.}
    \label{fig:chronos_attack}
\vspace{-10pt}
\end{figure}

\subsection{Predictability of the hourly DNS lookup}
While not explicitly stated in the Chronos proposal, the mechanism of querying the DNS hourly may give an attacker the possibility of knowing the DNS queries' timing, since it is done hourly for 24 hours. If the attacker can predict queries because they are only done at full-hour, or because he was able to observe previous queries, this facilitates off-path DNS poisoning by removing the necessity to trigger the query himself.

\subsection{Combining the results of all DNS responses}
The Chronos proposal queries the DNS multiple times to gather a big enough pool of servers. This queries are spread over 24 hours, mainly to prevent gathering the same NTP servers multiple times because of caching or round-robin mechanisms at the nameservers of {\footnotesize{\tt pool.ntp.org}}. However, Chronos does not make any efforts to prevent even a single malicious DNS response to influence the outcome of the whole pool-generation process, neither by checking the values of the response's Time-to-live value or the number of addresses given.

\subsection{DNS poisoning attack against Chronos}
Using the second of these two discovered weaknesses, we now present an enhanced DNS poisoning attack against Chronos which requires poisoning the DNS only once in this 24-hour period, shown in Figure \ref{fig:chronos_attack}. First we allow that the client has already done a number of $N$ of the 24 DNS requests before the attack starts, this could have been the result of the attacker failing to perform the attack a number of times. These queries where answered by the nameservers for {\footnotesize{\tt pool.ntp.org}}, which give 4 IP addresses per DNS request. This results in a number of $4N$ honest nameservers already in the generated server pool before the attack starts. Then, the attacker poisons the DNS resolver with as many addresses he can fit into a single DNS response (up to 89 for a single, non-fragmented UDP response) and sets the TTL to a value bigger than 24 hours, to cause any subsequent DNS requests to be answered from cache instead of the original nameserver. With this attack, the attacker adds 89 attacker-controlled IP addresses to the server pool and effectively ends the server pool-generation algorithm since subsequent DNS queries will always be answered with the cached, attacker-given records from the DNS resolver. The pool does now contain $4N$ honest and 89 malicious servers. To fulfil the requirement on a successful attack on Chronos, the attacker needs to control $\frac{2}{3}$ of the servers in the pool, so we get the maximum number for $N$ from $\frac{2}{3} \cdot (89 + 4N) \leq 89$ and conclude that the attacker succeeds with his attack if the cache poisoning succeeds before or during the 12th DNS transaction ($N<=11$). This means that the chances of a successful attack against Chronos are actually {\it higher} than against a traditional NTP client during boot-time, since the attacker effectively has 12 tries in 24 hours to succeed with the attack.

%% file: server-side.tex
\section{Measuring NTP Servers Attack Surface}
\label{sec:measurements}

\added{
In this section we measure the presence of server-side properties required for our attack to work: Rate-limiting support in NTP servers and fragmentation support in the NTP domain's nameservers.
}

\subsection{Rate Limiting of {\tt pool.ntp.org} NTP Servers}
\label{sec:ntpdns_ratelimit_upstream}

For successfully attacking NTP during run-time, the attacker needs to remove enough existing associations using NTP's rate-limiting mechanism. For this to work, the server actually needs to use rate-limiting. To find out how many servers do this, we conduct a study of all servers in {\it pool.ntp.org}.
We first query the {\it pool.ntp.org} namservers multiple times for all existing NTP country zones ({\it {\tt<}countrycode{\tt>}.pool.ntp.org}) and gather the union of all result IP addresses. This way we gathered a list of 2432 NTP servers. We then query each of those NTP servers 64 times, once per second and check for (1) KoD-packet from the NTP server or (2) the server stops responding after some queries, which indicates rate-limiting. To prevent false-positives because of packet loss as well as because some servers will answer a small fraction of queries, even during the client is rate-limited, we do this by taking the number of responses during the first and second half of the test and mark a server as rate-limiting if the first half has more than 8 additional responses compared to the second half. Using this method, we find that 780 (33\%) servers send KoD-messages and 904 (38\%) servers stopped sending responses during the second half of our test. We therefore conclude that around 38\% of servers in pool.ntp.org do rate-limiting, as sending KoD's is a clear indicator, but not every server sends a KoD message before rate-limiting the client.

\added{
\subsection{Fragmentation support of NTP Nameservers}
\label{sec:ntpdns_nameservers_3}

For DNS poisoning via IP fragment injection to work, nameservers have to support path MTU discovery (PMTUD) and fragment DNS responses on arrival of ICMP fragmentation needed messages. We tested this behaviour on nameservers of {\footnotesize{\tt pool.ntp.org}} and found that 16 out of 30 DNS nameservers fragment packets to fragments below 548 bytes on receiving ICMP fragmentation needed. None of those 30 namesevers supports DNSSEC for the {\footnotesize{\tt pool.ntp.org}} domain.

Furthermore, we evaluated PMTUD support of popular 1M domains including different NTP domains, not only those under {\footnotesize{\tt pool.ntp.org}}. In this measurement, we gathered data of 877,071 nameservers considering DNSSEC, fragmentation support and size of fragments emitted after arrival of ICMP fragmentation needed messages. We found that 7.66\% of domains do not support DNSSEC but emit fragmented DNS responses which makes them vulnerable to DNS cache-poisoning attacks via injection of IP fragments. Considering the minimum fragment size emitted by those domain's nameservers, see Fig.~\ref{fig:alexa_top_1m_domains_fragsize}, we found that most of those nameservers (83.2\%) fragment DNS responses down to a size of 548 bytes and 7.05\% even down to 292 bytes. This means that clients who use those NTP servers are vulnerable to the attack.
}
\added{
\begin{figure}[t]
    \centering
    \includegraphics[width=0.45\textwidth]{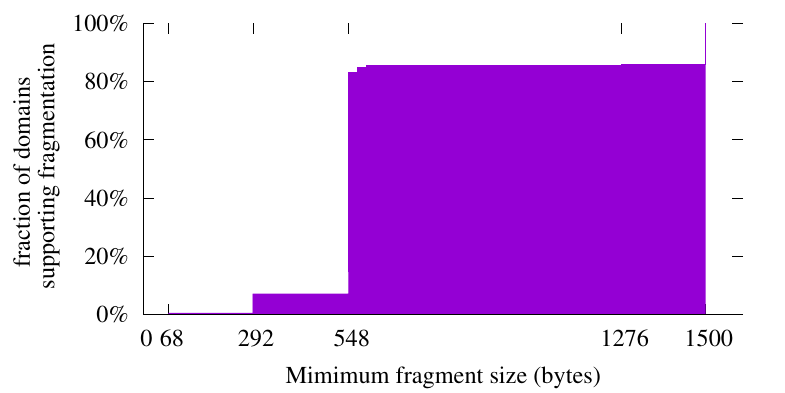}
    \caption{Cumulative distribution of fragment sizes emitted by popular 1M domains which do not support DNSSEC.} %
    \label{fig:alexa_top_1m_domains_fragsize}
    \vspace{-10pt}
\end{figure}

Whether the required response size for the DNS message to be fragmented is reached depends on the response generated by the nameserver, but attackers can employ several techniques to increase the size of a message.
For example, if the domain which is queried is (partially) under control of the attacker, attackers may trigger queries to exceptionally long sub-domains under the victim domain which is under attack. This is the case if a third-party system is used to trigger the DNS query as discussed in Section~\ref{sec:boottime}.
}

%% file: client-side.tex
\section{Measuring DNS Resolvers Attack Surface}
\label{sec:ntpdns_resolvers}
We measured DNS resolvers using an ad network and open DNS resolvers. In what follows we explain the study and report on our measurement results of vulnerable clients.

\subsection{Vulnerable Open DNS Resolvers}
To measure the client-side attack surface of our attack, we conduct a study of all Open DNS resolvers in the Censys DNS responder dataset \cite{censys15}. For attackers in a MitM position between DNS resolver and namserver, all clients are vulnerable to our attack, because the {\footnotesize{\tt pool.ntp.org}} namservers do not offer DNSSEC support. However, for off-path attackers without the possibility to get in path, e.g., by BGP-hijacking, attackers can still try to to fragmentation based DNS poisoning \cite{cns:frag:dns}. This requires the resolver to accept fragmented DNS responses, a property which we will explore here.

\subsubsection{\added{Finding resolvers used by NTP clients}}
\label{sec:openresolver-rdzero}

First, we need to find those open resolvers which are used by NTP clients. We do this by employing a technique which can detect if a DNS resolver has cached certain DNS records by sending DNS queries with the Recursion Desired (RD) Bit set to zero \cite{wills_inferring_nodate}. This is done with all records listed in Table \ref{tab:ntp_pool_cache_test}. We conclude that a resolver is used by NTP clients if any of those records is cached, since these domains generally serve no other purpose than NTP server discovery.

To ensure the cache testing method works as expected, for each resolver, we test if they respect the RD bit by first, querying a known non-cached domain and second, querying a known-cached domain which has been placed into the cache by a previously issued query with RD=1.

\begin{figure}[t]
    \centering
    \includegraphics[width=0.45\textwidth]{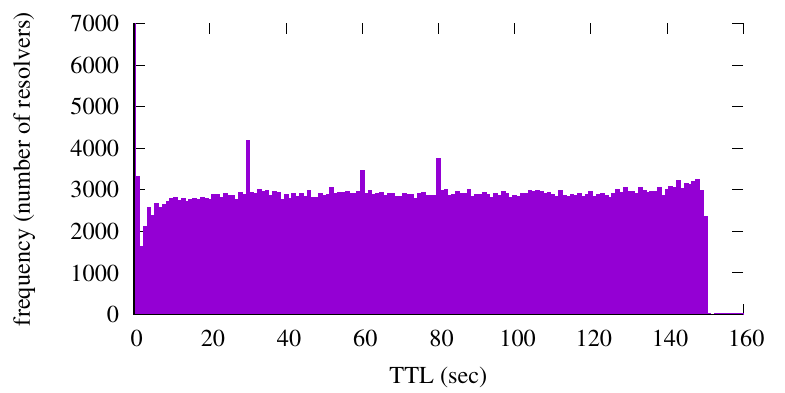}
    \caption{TTL values of cached NTP pool records in open resolvers}
    \label{fig:ntp_ttl_plot}
\vspace{-10pt}
\end{figure}

Using this method we probed 1,583,045 resolvers for cached records (1,674,103 resolvers in the dataset did not respond to any query), and verified the technique to be working 646,212 times. The results are shown in Table \ref{tab:ntp_pool_cache_test}. Out of the measured resolvers, 448,521 (69\%) resolvers had the \texttt{A} record for {\footnotesize{\tt pool.ntp.org}} cached. To further verify that our cache testing method indeed works, we provide TTL data for all these cached records in Figure \ref{fig:ntp_ttl_plot} and show that the values are uniformly distributed, as one would expect for cached records.

\subsubsection{\added{Measuring fragmentation support}}

We then perform a generic scan of open resolvers for fragmentation acceptance by querying a purposely created domain we control, where the nameserver always responds to DNS requests with fragmented packets, even if the size is way below the maximum MTU of the path to the tested resolver. Out of the resolvers which had any of the {\footnotesize{\tt pool.ntp.org}} records cached, 32\% accepted fragmented DNS responses and 68\% did not, which is roughly the same distribution as over the whole population of open resolvers (31\% fragmented DNS response acceptance).

\begin{table}[t]
    \centering
    \scriptsize
    \caption{{\footnotesize{\tt pool.ntp.org}} caching state in tested open resolvers}
    \label{tab:ntp_pool_cache_test}
    \begin{tabular}{|l|c|r|r|}
    \hline
                             & Cached  & \multicolumn{2}{c|}{Absolute Resolvers} \\
    \cline{3-4}
        Query                & in      & Cached & Not Cached \\
    \hline
        \texttt{pool.ntp.org IN NS}    & 58.28\% & 376,594 & 269,618 \\
        \texttt{pool.ntp.org IN A}     & 69.41\% & 448,521 & 197,691 \\
        \texttt{0.pool.ntp.org IN A}   & 63.92\% & 413,046 & 233,166 \\
        \texttt{1.pool.ntp.org IN A}   & 61.28\% & 395,971 & 250,241 \\
        \texttt{2.pool.ntp.org IN A}   & 61.55\% & 397,751 & 248,461 \\
        \texttt{3.pool.ntp.org IN A}   & 58.58\% & 378,525 & 267,687 \\ 
    \hline
    \end{tabular}
\end{table}

\subsection{Measurement via Ad-Network of Vulnerable DNS Resolvers}

We conducted a study of resolvers used by Web clients in the Internet, as many Web clients will also run NTP clients. For this purpose, we created an automated test website which was served by an advertisement network. This study is designed to test fragmentation support of those resolvers which is a requirement for off-path DNS-poisoning.

\subsubsection{Testing fragmentation support through Javascript}

To conduct the study, we used an ad network which serves our test website as an `popunder'-ad. This causes our test website to be loaded in a background browser window or tab. The test website then uses a script to load a series of images from different tests such that each image is to verify the successful DNS lookup of its corresponding domain, similar to the study done by Lian et al. in 2013 to identify DNSSEC deployment \cite{lian2013measuring}. We conducted the following tests where {\it T} represents a randomly generated token to differentiate clients and avoid caching. Moreover, our nameserver fragmented the responses irrespective of any path-MTU-discovery results:
\begin{itemize}
\small
\item {\footnotesize{\tt T.baseline.domain.com}} - normal A record
\item {\footnotesize{\tt T.ftiny.domain.com}} - response fragmented to 68 bytes
\item {\footnotesize{\tt T.fsmall.domain.com}} - response fragmented to 296 bytes
\item {\footnotesize{\tt T.fmedium.domain.com}} - response fragmented to 580 bytes
\item {\footnotesize{\tt T.fbig.domain.com}} - response fragmented to 1280 bytes
\item {\footnotesize{\tt sigfail.domain.com}} - incorrectly DNSSEC-signed record
\item {\footnotesize{\tt sigright.domain.com}} - correctly DNSSEC-signed record
\end{itemize}
We used a customised nameserver for our domain to guarantee that responses for queries about the \{ftiny, fsmall, fmedium, fbig\} domains are always in at least two fragments of their corresponding MTU size. After the test is finished, the javascript records if the images are loaded via onsucess()/onerror() event handlers and finally push back the results to the server. To filter out invalid results, we removed any results where the client had the page open for less than 30 seconds, to prevent errors because of timeouts at the resolver level. We also remove results which failed the {\it baseline} or {\it sigright tests} as this tests are designed as baseline tests to verify that the testing method works.

\subsubsection{Results}

In our first run of the study we published our test website using the ad network without targeting any specific regions. In this run, which we call dataset 1, we got valid results from 5847 unique clients. Because our first run gave only 120 results for clients in the North America region, we conducted a second run
only targeting clients in the United States and Canada, which we call dataset 2.

Results for both datasets are displayed in Table \ref{tab:ad_network_results}, which group clients by region
as well as device type (determined using the HTTP user-agent header). Furthermore, by correlating the randomised
{\it T} values with nameserver logs, we figured that 791 clients uses resolver operated by Google LLC, all of which
filter fragments of sizes below `big'. The column `Accepts any fragment size' shows
results of clients which used resolvers accepting at least one fragmented response. Depending on the attacker's ability to trigger a large enough fragmented response at the nameserver, these clients are vulnerable to off-path de-fragmentation DNS poisoning.

Fragment acceptance does not change dramatically with the fragment size, 
many resolvers (64\%) accept even the tiniest possible fragment size with MTU=68, this number increases to 77\% for medium fragments and 86\% for big fragments. This distribution across various fragment sizes behaves similar in for all regions and device groups.
DNSSEC validation ranges between 19.14\% and 28.94\%, a meaningful increase over the 3\% measured by Lian et al. in 2013 \cite{lian2013measuring}.

\added{
\subsubsection{Finding shared DNS resolvers}
\label{sec:sharedresolvers}\label{sc:timing_cache_test}
The attacker can utilise different systems for initiating the cache poisoning attack and injecting a malicious record into the victim DNS resolver's cache, and is not limited to attacking the query issued by the victim NTP client. Popular systems present in almost any networks are Email servers or web clients/proxies; for instance, Email issues queries to DNS resolver upon receipt of new Emails for performing domain-based anti-spam validation. We perform an Internet scale measurement for resolvers which are used by Email and web as well as by NTP clients. Then we cause the queries via Email or web clients for attacking NTP.

For web clients we leverage an ad-network causing the clients to issue queries through their DNS resolvers to our nameserver and for Email we send Email messages to the measured domains, which in turn cause the DNS resolvers to issue queries to our nameservers.
First, we sent direct DNS queries to all resolvers to figure out if they are perhaps open resolvers. Secondly, we conduct
a small portscan for SMTP servers in the /24 networks of all resolvers and web clients to find SMTP servers share the same DNS resolver. We found 6416 SMTP servers for all of which we sent test-emails causing bounces to our own nameserver, which lets us determine the DNS resolver used by the SMTP server.
Finally we overlap the list of resolvers used by SMTP servers with the list of resolvers used by ad clients. Out of 18,668 resolvers used by web clients:
\begin{itemize}
\small
\item 16,088 (86.2\%) are only used by web clients,
\item 2,116 (11.3\%) are used by web clients and SMTP servers,
\item 426 (2.3\%) are open resolvers,
\item 38 (0.2\%) are open and used by both web and SMTP.
\end{itemize}
Combining SMTP and open resolver results, we find that an attacker could trigger queries for at least 2580 (13.8\%)
of resolvers using either SMTP or direct queries. This should be seen as a lower bound, because our very imperfect
method of finding the SMTP servers used to trigger the queries. If such servers exist, but are located outside of
the /24 network ranges we scanned, they are not considered in this study.
}

\begin{table}[t]
    \centering
    \scriptsize

    \caption{Results of client resolver study using Ads}
    \label{tab:ad_network_results}
\begin{tabular}{|l|rr|rr|r|r|}
\hline

&\multicolumn{2}{c|}{Accepts tiny}    &\multicolumn{2}{c|}{Accepts any}    &total&data\\
&\multicolumn{2}{c|}{fragments (68 B)}&\multicolumn{2}{c|}{fragment size}  &     &set\\
\hline
\hline
Asia&1845&58.22\%&2863&90.34\%&3169&1 \\
\hline
Africa&222&73.27\%&290&95.71\%&303&1 \\
\hline
Europe&1010&72.66\%&1277&91.87\%&1390&1 \\
\hline
Northern America&1352&58.43\%&1757&75.93\%&2314&2 \\
\hline
Latin America&572&68.26\%&759&90.57\%&838&1 \\
\hline
\hline
ALL&3742&64.00\%&5320&90.99\%&5847&1 \\
\hline
\hline
Without Google&3439&68.02\%&4555&90.09\%&5056&1 \\
\hline
PC&1664&60.80\%&2447&89.40\%&2737&1 \\
\hline
Mobile,Tablet&2077&66.83\%&2871&92.37\%&3108&1 \\
\hline

\end{tabular}

\vspace{-15pt}
\end{table}

\added{
To find web clients' resolvers who also serve NTP clients, we develop another an additional cache-testing method.}
This method works by using the timing side-channel which is exposed by the latency of a query: If a record is not cached by the resolver, subsequent queries for the same domain should be faster then the first query because they can be answered from cache instead of asking the remote nameserver(s). This can be automatically tested by comparing the latency difference $t_{first}$ of the first query and the average $t_{avg}$ of subsequent queries to a threshold $T$ and call the record cached, if $t_{first} - t_{avg} < T$. However, because we cannot control other variables which have an impact of the query latency, such as RTT to the tested resolver and the fact that some resolvers may have NS-records for parent zones like $ntp.org$ cached and others have not, there is no obvious value to choose $T$.

\added{
To find out if the method works, we executed it on the open resolver dataset first. We find that there is no obviously spot-able distribution of the timing latency $t_{first} - t_{avg}$ into two groups (cached and non-cached) and therefore no way to reasonably choose a value for $T$. } %

\added{
We conduct that executing such a test successfully requires further evaluation and multiple test passes including eviction of the cached records to control the variances in round-trip-time and other factors influencing the timing of the query. As this may substantially reduce the performance of affected resolvers and Web clients, we resigned from testing caching status of {\footnotesize{\tt pool.ntp.org}} against web clients' resolvers and instead assume that many resolvers in this dataset are indeed used by NTP clients, as NTP clients are part of most modern operating systems.
}

\begin{figure}
    \centering
    \includegraphics[width=0.45\textwidth]{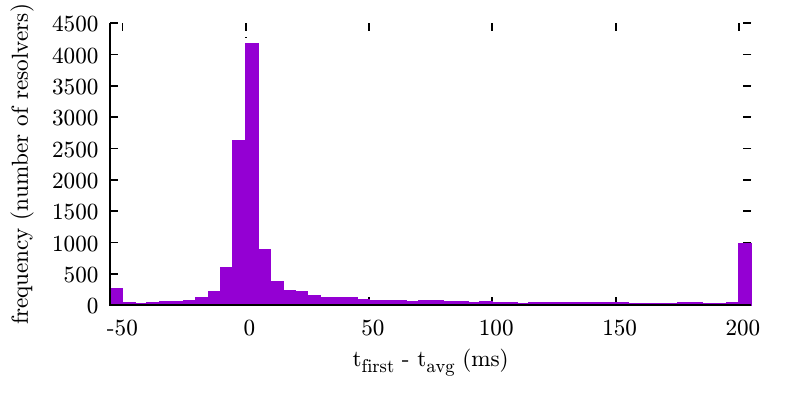}
    \caption{Distribution of latency difference when querying open resolvers for \texttt{pool.ntp.org IN NS}. Values below -50 ms and above 200 ms are summed up on the sides.}%
    \label{fig:ntp_timing_test_distribution}
\vspace{-10pt}
\end{figure}

%% file: mitigations.tex
\section{Countermeasures and Mitigations}
\label{sec:mitigations}
In this section, we discuss the state of security of DNS and NTP. Essentially currently both are vulnerable and the dependency of NTP on DNS makes our off-path attacks possible. As an immediate countermeasure we recommend not to use DNS for NTP and instead to use a list of static IP addresses for NTP servers or to deploy distributed proposals like \cite{jeitner:consensus:dsn:20}.

{\bf DNS Security.} DNSSEC could make the cache poisoning attacks practically impossible. However, studies showed less than 10\% of the DNS resolvers validate DNS responses, \cite{herzberg2013towards,fukuda2013technique,lian2013measuring}. Our own validation with Ad-Net showed that DNSSEC validation depends on the geo-location and ranges between 19.14\% and 28.94\%. On the other hand, only about 1\% of the domains are signed with DNSSEC and in our measurement we found {\em only one} signed NTP domain {\tt time.cloudflare.com}, so even if the resolvers performed strict validation this would currently not help. Furthermore, recent studies found problems in DNSSEC keys generation making many signed domains vulnerable despite DNSSEC, \cite{dai2016dnssec,shulman:nsdi17,chung2017longitudinal}. 

{\bf NTP Security.} Different recommendations for securing NTP were proposed, such as TLS for NTP (NTS) \cite{I-D.ietf-ntp-using-nts-for-ntp}, or Roughtime \cite{I-D.roughtime-aanchal} with proof of misbehaviour (wrong time) to report bad servers. However, none is deployed or used, mostly due to the significant changes to the NTP ecosystem that they require, or assumptions which cannot be fulfilled in practice.

%% file: conclusions.tex
\section{Conclusions}
\label{sec:conclusion}

Our attacks demonstrate that the NTP ecosystem is still vulnerable to off-path attacks. In this work we improve over the attacks in \cite{malhotra_attacking_2016} by demonstrating more effective and practical attacks. We show that, as long as DNS is insecure, our attacks apply even to the security enhanced NTP client with Chronos \cite{ntp:chronos}. \added{Launching off-path time-shifting attacks against NTP is challenging: (1) the attacker must trigger a query or be able to predict the timing when the query is triggered and (2) once poisoning is successful, the attacker needs to cause the NTP client to use the new malicious mapping. We developed techniques which allow to handle both these challenges and performed Internet wide measurements to show the applicability of our methodologies.
}

Although the vulnerabilities that we exploit are not in NTP, our attacks demonstrate the risks of building security on insecure Internet foundations, and the risks of analysing security in isolated or ideal environment which does not reflect the situation on the real Internet.

%% file: main.bbl
\begin{thebibliography}{10}
\providecommand{\url}[1]{#1}
\csname url@samestyle\endcsname
\providecommand{\newblock}{\relax}
\providecommand{\bibinfo}[2]{#2}
\providecommand{\BIBentrySTDinterwordspacing}{\spaceskip=0pt\relax}
\providecommand{\BIBentryALTinterwordstretchfactor}{4}
\providecommand{\BIBentryALTinterwordspacing}{\spaceskip=\fontdimen2\font plus
\BIBentryALTinterwordstretchfactor\fontdimen3\font minus
  \fontdimen4\font\relax}
\providecommand{\BIBforeignlanguage}[2]{{%
\expandafter\ifx\csname l@#1\endcsname\relax
\typeout{** WARNING: IEEEtran.bst: No hyphenation pattern has been}%
\typeout{** loaded for the language `#1'. Using the pattern for}%
\typeout{** the default language instead.}%
\else
\language=\csname l@#1\endcsname
\fi
#2}}
\providecommand{\BIBdecl}{\relax}
\BIBdecl

\bibitem{malhotra_attacking_2016}
\BIBentryALTinterwordspacing
A.~Malhotra and S.~Goldberg, ``\BIBforeignlanguage{en}{Attacking {NTP}'s
  {Authenticated} {Broadcast} {Mode}},'' \emph{\BIBforeignlanguage{en}{ACM
  SIGCOMM Computer Communication Review}}, vol.~46, no.~1, pp. 12--17, May
  2016. [Online]. Available:
  \url{http://dl.acm.org/citation.cfm?doid=2935634.2935637}
\BIBentrySTDinterwordspacing

\bibitem{ntp:chronos}
O.~Deutsch, N.~R. Schiff, D.~Dolev, and M.~Schapira, ``Preventing ({Network})
  {Time} {Travel} with {Chronos},'' in \emph{Proceedings 2018 {Network} and
  {Distributed} {System} {Security} {Symposium}}, San Diego, CA, 2018.

\bibitem{ntp:draft}
\BIBentryALTinterwordspacing
N.~S. Rozen, D.~Dolev, T.~Mizrahi, and M.~Schapira,
  ``\BIBforeignlanguage{en}{{A Secure Selection and Filtering Mechanism for the
  Network Time Protocol}}.'' [Online]. Available:
  \url{https://tools.ietf.org/html/draft-schiff-ntp-chronos-03}
\BIBentrySTDinterwordspacing

\bibitem{DBLP:conf/sigcomm/BallaniFZ07}
H.~Ballani, P.~Francis, and X.~Zhang, ``A study of prefix hijacking and
  interception in the internet,'' in \emph{{SIGCOMM}}.\hskip 1em plus 0.5em
  minus 0.4em\relax {ACM}, 2007, pp. 265--276.

\bibitem{DBLP:journals/pieee/ButlerFMR10}
K.~R.~B. Butler, T.~R. Farley, P.~D. McDaniel, and J.~Rexford, ``A survey of
  {BGP} security issues and solutions,'' \emph{Proceedings of the {IEEE}},
  vol.~98, no.~1, pp. 100--122, 2010.

\bibitem{DBLP:journals/comsur/HustonRA11}
G.~Huston, M.~Rossi, and G.~J. Armitage, ``Securing {BGP} - {A} literature
  survey,'' \emph{{IEEE} Communications Surveys and Tutorials}, vol.~13, no.~2,
  pp. 199--222, 2011.

\bibitem{dyn08}
{DYN}, ``Pakistan hijacks youtube,'' 24 February 2008,
  \url{https://dyn.com/blog/pakistan-hijacks-youtube-1/}.

\bibitem{Cowie10}
J.~Cowie, ``China’s 18-minute mystery,'' 18 November 2010,
  \url{https://dyn.com/blog/chinas-18-minute-mystery/}.

\bibitem{cns:frag:dns}
A.~Herzberg and H.~Shulman, ``Fragmentation {C}onsidered {P}oisonous: or
  one-domain-to-rule-them-all.org,'' in \emph{IEEE CNS 2013. The Conference on
  Communications and Network Security, Washington, D.C., U.S.}\hskip 1em plus
  0.5em minus 0.4em\relax IEEE, October 2013.

\bibitem{czyz2014taming}
J.~Czyz, M.~Kallitsis, M.~Gharaibeh, C.~Papadopoulos, M.~Bailey, and M.~Karir,
  ``Taming the 800 pound gorilla: The rise and decline of ntp ddos attacks,''
  in \emph{Proceedings of the 2014 Conference on Internet Measurement
  Conference}.\hskip 1em plus 0.5em minus 0.4em\relax ACM, 2014, pp. 435--448.

\bibitem{Amplification:Hell}
C.~Rossow, ``{Amplification Hell: Revisiting Network Protocols for DDoS
  Abuse},'' in \emph{Proceedings of the Network and Distributed System Security
  (NDSS) Symposium}, February 2014.

\bibitem{ntp:kiayias_security_2017}
A.~Malhotra, M.~Van~Gundy, M.~Varia, H.~Kennedy, J.~Gardner, and S.~Goldberg,
  ``\BIBforeignlanguage{en}{The {Security} of {NTP}’s {Datagram}
  {Protocol}},'' in \emph{\BIBforeignlanguage{en}{Financial {Cryptography} and
  {Data} {Security}}}, A.~Kiayias, Ed.\hskip 1em plus 0.5em minus 0.4em\relax
  Cham: Springer International Publishing, 2017, vol. 10322, pp. 405--423.

\bibitem{mills2016computer}
D.~L. Mills, \emph{Computer network time synchronization: the network time
  protocol on earth and in space}.\hskip 1em plus 0.5em minus 0.4em\relax CRC
  press, 2016.

\bibitem{rfc7384}
\BIBentryALTinterwordspacing
T.~Mizrahi, ``Security requirements of time protocols in packet switched
  networks,'' Internet Requests for Comments, RFC Editor, RFC 7384, October
  2014. [Online]. Available: \url{http://www.rfc-editor.org/rfc/rfc7384.txt}
\BIBentrySTDinterwordspacing

\bibitem{selvi2014bypassing}
J.~Selvi, ``Bypassing http strict transport security,'' \emph{Black Hat
  Europe}, 2014.

\bibitem{selvi2015breaking}
------, ``Breaking ssl using time synchronisation attacks,'' \emph{DEF CON
  Hacking Conference}, 2015.

\bibitem{malhotra_attacking_2016-1}
\BIBentryALTinterwordspacing
A.~Malhotra, I.~E. Cohen, E.~Brakke, and S.~Goldberg,
  ``\BIBforeignlanguage{en}{Attacking the {Network} {Time} {Protocol}},'' in
  \emph{\BIBforeignlanguage{en}{Proceedings 2016 {Network} and {Distributed}
  {System} {Security} {Symposium}}}.\hskip 1em plus 0.5em minus 0.4em\relax San
  Diego, CA: Internet Society, 2016. [Online]. Available:
  \url{https://www.ndss-symposium.org/wp-content/uploads/sites/25/2017/09/attacking-network-time-protocol.pdf}
\BIBentrySTDinterwordspacing

\bibitem{dowling2016authenticated}
B.~Dowling, D.~Stebila, and G.~Zaverucha, ``Authenticated network time
  synchronization,'' in \emph{25th {USENIX} Security Symposium ({USENIX}
  Security 16)}, 2016, pp. 823--840.

\bibitem{mizrahi2012slave}
T.~Mizrahi, ``Slave diversity: Using multiple paths to improve the accuracy of
  clock synchronization protocols,'' in \emph{2012 IEEE International Symposium
  on Precision Clock Synchronization for Measurement, Control and Communication
  Proceedings}.\hskip 1em plus 0.5em minus 0.4em\relax IEEE, 2012, pp. 1--6.

\bibitem{shpiner2013multi}
A.~Shpiner, Y.~Revah, and T.~Mizrahi, ``Multi-path time protocols,'' in
  \emph{2013 IEEE International Symposium on Precision Clock Synchronization
  for Measurement, Control and Communication (ISPCS) Proceedings}.\hskip 1em
  plus 0.5em minus 0.4em\relax IEEE, 2013, pp. 1--6.

\bibitem{Vixie95}
P.~Vixie, ``{DNS} and {BIND} security issues,'' in \emph{Proceedings of the 5th
  Symposium on {UNIX} Security}.\hskip 1em plus 0.5em minus 0.4em\relax
  Berkeley, CA, USA: USENIX Association, jun 1995, pp. 209--216.

\bibitem{Bernstein:DNS}
D.~J. Bernstein, ``{DNS Forgery},'' Internet publication at
  http://cr.yp.to/djbdns/forgery.html, November 2002.

\bibitem{kaminsky:dns}
D.~Kaminsky, ``{I}t's the {E}nd of the {C}ache {A}s {W}e {K}now {I}t,'' in
  \emph{Black Hat conference}, August 2008,
  \url{http://www.blackhat.com/presentations/bh-jp-08/bh-jp-08-Kaminsky/BlackHat-Japan-08-Kaminsky-DNS08-BlackOps.pdf}.

\bibitem{hubert2009measures}
A.~Hubert and R.~Van~Mook, ``Measures for making dns more resilient against
  forged answers,'' RFC 5452, January, Tech. Rep., 2009.

\bibitem{gilad2014off}
Y.~Gilad and A.~Herzberg, ``Off-path tcp injection attacks,'' \emph{ACM
  Transactions on Information and System Security (TISSEC)}, vol.~16, no.~4,
  p.~13, 2014.

\bibitem{shulman2015towards}
H.~Shulman and M.~Waidner, ``Towards security of internet naming
  infrastructure,'' in \emph{European Symposium on Research in Computer
  Security}.\hskip 1em plus 0.5em minus 0.4em\relax Springer, 2015, pp. 3--22.

\bibitem{brandt2018domain}
M.~Brandt, T.~Dai, A.~Klein, H.~Shulman, and M.~Waidner, ``{Domain Validation++
  For MitM-Resilient PKI},'' in \emph{Proceedings of the 2018 ACM SIGSAC
  Conference on Computer and Communications Security}.\hskip 1em plus 0.5em
  minus 0.4em\relax ACM, 2018, pp. 2060--2076.

\bibitem{herzberg2012antidotes}
A.~Herzberg and H.~Shulman, ``Antidotes for dns poisoning by off-path
  adversaries,'' in \emph{2012 Seventh International Conference on
  Availability, Reliability and Security}.\hskip 1em plus 0.5em minus
  0.4em\relax IEEE, 2012, pp. 262--267.

\bibitem{frag:vulnerable:journal}
Y.~Gilad and A.~Herzberg, ``{Fragmentation Considered Vulnerable},''
  \emph{{ACM} Transactions on Information and System Security {(TISSEC)}},
  vol.~15, no.~4, pp. 16:1--16:31, April 2013, a preliminary version appeared
  in WOOT 2011.

\bibitem{rytilahti_masters_2018}
T.~Rytilahti, D.~Tatang, J.~Köpper, and T.~Holz, ``Masters of {Time}: {An}
  {Overview} of the {NTP} {Ecosystem},'' in \emph{2018 {IEEE} {European}
  {Symposium} on {Security} and {Privacy} ({EuroS} {P})}, Apr. 2018, pp.
  122--136.

\bibitem{I-D.schiff-ntp-chronos}
\BIBentryALTinterwordspacing
N.~Schiff, D.~Dolev, T.~Mizrahi, and M.~Schapira, ``A secure selection and
  filtering mechanism for the network time protocol version 4,'' Working Draft,
  IETF Secretariat, Internet-Draft draft-schiff-ntp-chronos-03, September 2019.
  [Online]. Available:
  \url{http://www.ietf.org/internet-drafts/draft-schiff-ntp-chronos-03.txt}
\BIBentrySTDinterwordspacing

\bibitem{censys15}
Z.~Durumeric, D.~Adrian, A.~Mirian, M.~Bailey, and J.~A. Halderman, ``A search
  engine backed by {I}nternet-wide scanning,'' in \emph{22nd {ACM} Conference
  on Computer and Communications Security}, Oct. 2015.

\bibitem{wills_inferring_nodate}
\BIBentryALTinterwordspacing
C.~E. Wills, M.~Mikhailov, and H.~Shang, ``Inferring relative popularity of
  internet applications by actively querying dns caches,'' in \emph{Proceedings
  of the 3rd ACM SIGCOMM Conference on Internet Measurement}, ser. IMC
  ’03.\hskip 1em plus 0.5em minus 0.4em\relax New York, NY, USA: Association
  for Computing Machinery, 2003, p. 78–90. [Online]. Available:
  \url{https://doi.org/10.1145/948205.948216}
\BIBentrySTDinterwordspacing

\bibitem{lian2013measuring}
W.~Lian, E.~Rescorla, H.~Shacham, and S.~Savage, ``{Measuring the Practical
  Impact of DNSSEC Deployment},'' in \emph{Proceedings of USENIX Security},
  2013.

\bibitem{jeitner:consensus:dsn:20}
P.~Jeitner, H.~Shulman, and M.~Waidner, ``{Secure Consensus Generation with
  Distributed DoH},'' in \emph{50th Annual {IEEE/IFIP} International Conference
  on Dependable Systems and Networks Workshops, {DSN} Workshops 2020, Valencia,
  Spain, June, 2020}.

\bibitem{herzberg2013towards}
A.~Herzberg and H.~Shulman, ``Towards adoption of dnssec: Availability and
  security challenges.'' \emph{IACR Cryptol. ePrint Arch.}, vol. 2013, p. 254,
  2013.

\bibitem{fukuda2013technique}
K.~Fukuda, S.~Sato, and T.~Mitamura, ``A technique for counting dnssec
  validators,'' in \emph{2013 Proceedings IEEE INFOCOM}.\hskip 1em plus 0.5em
  minus 0.4em\relax IEEE, 2013, pp. 80--84.

\bibitem{dai2016dnssec}
T.~Dai, H.~Shulman, and M.~Waidner, ``Dnssec misconfigurations in popular
  domains,'' in \emph{International Conference on Cryptology and Network
  Security}.\hskip 1em plus 0.5em minus 0.4em\relax Springer, 2016, pp.
  651--660.

\bibitem{shulman:nsdi17}
H.~Shulman and M.~Waidner, ``{One Key to Sign Them All Considered Vulnerable:
  Evaluation of DNSSEC in Signed Domains},'' in \emph{The 14th USENIX Symposium
  on Networked Systems Design and Implementation (NSDI)}.\hskip 1em plus 0.5em
  minus 0.4em\relax USENIX, 2017.

\bibitem{chung2017longitudinal}
T.~Chung, R.~van Rijswijk-Deij, B.~Chandrasekaran, D.~Choffnes, D.~Levin, B.~M.
  Maggs, A.~Mislove, and C.~Wilson, ``A longitudinal, end-to-end view of the
  dnssec ecosystem,'' in \emph{USENIX Security}, 2017.

\bibitem{I-D.ietf-ntp-using-nts-for-ntp}
\BIBentryALTinterwordspacing
D.~Franke, D.~Sibold, K.~Teichel, M.~Dansarie, and R.~Sundblad, ``Network time
  security for the network time protocol,'' Working Draft, IETF Secretariat,
  Internet-Draft draft-ietf-ntp-using-nts-for-ntp-20, July 2019. [Online].
  Available:
  \url{http://www.ietf.org/internet-drafts/draft-ietf-ntp-using-nts-for-ntp-20.txt}
\BIBentrySTDinterwordspacing

\bibitem{I-D.roughtime-aanchal}
\BIBentryALTinterwordspacing
A.~Malhotra, A.~Langley, and W.~Ladd, ``Roughtime,'' Working Draft, IETF
  Secretariat, Internet-Draft draft-roughtime-aanchal-03, July 2019. [Online].
  Available:
  \url{http://www.ietf.org/internet-drafts/draft-roughtime-aanchal-03.txt}
\BIBentrySTDinterwordspacing

\end{thebibliography}
